\documentclass[prl,twocolumn]{revtex4}
\usepackage{array,amsmath,amssymb,graphicx}
\begin{document}

\noindent {\bf Comment on ``Analytic Structure of One-Dimensional
Localization Theory: Re-Examining Mott's Law''}

\vspace{12pt}

The low-frequency conductivity $\sigma(\omega)$ of a disordered Fermi
gas in one spatial dimension is governed by the Mott--Berezinskii
law~\cite{Mott_68,Berezinskii_73,Comment_on_log}
\begin{equation}
\sigma(\omega) = \sigma_0 \{2 \zeta(3) \varepsilon - \varepsilon^2
[\ln^2 \varepsilon + (2 \mathbb{C} - 3)\ln \varepsilon - c]\},
\label{sigma_Mott}
\end{equation}
Here $\sigma_0$ is the Drude conductivity, $\varepsilon = -2 i\omega \tau$ is
a small parameter in the problem, $\tau$ is the elastic scattering
time, $\mathbb{C}$ is the Euler constant, and $c = O(1)$ is another
constant. In a recent Letter~\cite{Gogolin_00} Gogolin claimed that
the expansion in powers of $\ln\varepsilon$ inside the square brackets in
Eq.~(\ref{sigma_Mott}) starts with $\ln^3\varepsilon$ term, challenging our
basic ideas about disordered systems~\cite{Mott_68}, as well as
previous analytical~\cite{Berezinskii_73,Abrikosov_77,Gorkov_83} and
numerical~\cite{Gogolin_78} work.
Below we reinstate the validity of Mott--Berezinskii law by pointing out
two calculational errors in Gogolin's paper. We also present
numerical results for $\sigma$, which fully agree with Eq.~(\ref{sigma_Mott}).

Gogolin expresses $\sigma(\omega)$ as a sum over certain matrix
elements $f_n$ of the eigenfunctions of Berezinskii's second-order
differential equation:
\begin{equation}
\sigma(\omega) = \sum\limits_{n = 1}^\infty f_n,
\quad
f_n = \frac{\pi^2 \sigma_0 \varepsilon}{\cosh^3 \pi k_n}
\frac{B^2_{n\varepsilon}}{k_n^2 + \frac14}.
\label{sigma_sum_f}
\end{equation}
The wavenumbers $k_n$ are to be determined from the
equation $g(k_n) = 2 \pi n$, where
\begin{equation}
g(k) = 2 k {\cal L} - i \ln\left[
\frac{\Gamma^2(1 + i k) \Gamma(\frac12 - i k)}
     {\Gamma^2(1 - i k) \Gamma(\frac12 + i k)}\right]
\label{g} 
\end{equation}
and ${\cal L} = \ln (16/\varepsilon)$.  The Poisson summation formula
brings Eq.~(\ref{sigma_sum_f}) to the form
\begin{equation}
\sigma(\omega) = \frac{1}{4\pi}
\sum\limits_{p = -\infty}^\infty \int d k f(k) e^{i p g(k)} g^\prime(k).
\label{Poisson} 
\end{equation}
The third power of $\ln\varepsilon$ was obtained in Ref.~\cite{Gogolin_00}
because $g(k) = 2 k {\cal L}$ was used instead of the full
expression~(\ref{g}). This is the crucial mistake in that paper.

To get the correct coefficient in front of the subleading term in
Eq.~(\ref{sigma_Mott}), one more mistake has to be rectified.  Namely,
one has to use the following expression for the normalization factor
$B_{n\varepsilon}$:
\begin{equation}
\frac{1}{B_{n\varepsilon}^2} =
\frac{\pi {\cal L}}{2 k_n \sinh \pi k_n}
- \frac{\pi}{4}
\frac{\Gamma(\frac14 + \frac{i k_n}{2})\Gamma(\frac14 - \frac{i k_n}{2})}
     {\Gamma(\frac34 + \frac{i k_n}{2})\Gamma(\frac34 - \frac{i k_n}{2})}.
\label{B} 
\end{equation}
Gogolin retains only the first term, presumably because it contains
the large logarithm ${\cal L}$.  However, the second term diverges at
the point $k_n = i / 2$ where the integrand in Eq.~(\ref{Poisson}) has
a pole. Therefore, this term must be retained.  Substituting more
accurate expressions~(\ref{g}) and (\ref{B}) into
Eqs.~(\ref{sigma_sum_f}) and (\ref{Poisson}), we recover
Eq.~(\ref{sigma_Mott}).

\begin{figure}
\includegraphics[width=2.2in,angle=-90,bb=12 12 576 693]{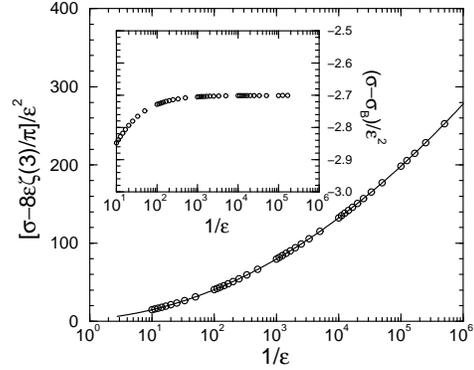}
\vspace{0.1in}
\setlength{\columnwidth}{3.2in}
\caption{
Numerical results for $\sigma$ (see text).
\label{Numerics}
}
\end{figure}

As a final check, we solved Berezinskii's recursive
equations~\cite{Berezinskii_73,Gogolin_00} numerically for small real
$\varepsilon$, slightly refining the algorithm described in
Ref.~\cite{Gogolin_78}.
The solid line in Figure~\ref{Numerics} is the best fit of $\sigma(\varepsilon)$
to the form $\sigma_B(\varepsilon) = 8 \zeta(3) \varepsilon / \pi -
\beta\varepsilon^2 (\ln^2\varepsilon - \gamma\ln\varepsilon + \delta)$ ($\sigma_0
= 4 / \pi$ in our system of units), which is provided by
$\beta=1.272(5)$, $\gamma=1.85(1)$, and $\delta=2.11(1)$, in excellent
agreement with $\beta = 4 / \pi=1.2732\ldots$, $\gamma = 3 - 2\mathbb{C}
= 1.8456\ldots$ predicted by Eq.~(\ref{sigma_Mott}). In the inset, we
plot $[\sigma(\varepsilon) - \sigma_B] / \varepsilon^2$ using the latter
values of $\beta$ and $\gamma$ and $\delta = 0$. This quantity tends to
a constant at small $\varepsilon$ in accordance with Eq.~(\ref{sigma_Mott}).

%
%

{\it Acknowledgements\/}. M.~M.~F. is supported by Pappalardo
Fellowship in Physics at MIT and Z.~W. by DOE Grant No. DE-FG02-99ER45747.

\vspace{12pt}
\vbox{
\indent Michael M. Fogler and Ziqiang Wang$^\dagger$\\
\indent Department of Physics\\
\indent Massachusetts Institute of Technology\\
\indent 77 Massachusetts Avenue\\
\indent Cambridge, MA 02139

\vspace{6pt}

\indent $^\dagger$On subbatical leave from\\
\indent Department of Physics\\
\indent Boston College\\
\indent Chestnut Hill, MA 02467
}

\pacs{\mbox{\hspace{-2cm} PACS numbers: 72.15.Rn, 73.20.Fz}}

\end{document}